\documentclass[a4paper,11pt]{article}
\usepackage{enumitem}
\usepackage{setspace}
\usepackage{pos}
\usepackage{subfig}
\usepackage{lineno}
\usepackage{cancel}
\usepackage{natbib}

\title{CRPropa 3.3: Toward a Unified Multi-Messenger Framework from GeV to ZeV Energies}

\ShortTitle{CRPropa 3.3: Toward a Unified Multi-Messenger Framework}

\author[1, 2]{S.~Aerdker}
\author[3, 4]{R.~{Alves Batista}}
\author[1, 2, 5]{J.~Becker Tjus}
\author[6, 7]{G.~Di Marco}
\author[1, 2]{J.~Dörner}
\author[8, 2]{K.-H.~Kampert}
\author[1, 2]{L.~Merten}
\author[8]{L.~Morejon}
\author[]{G.~Müller}
\author[9]{P.~Reichherzer}
\author*[10,11,12]{A.~Saveliev}
\emailAdd{crpropa@desy.de}
\author[1, 2]{L.~Schlegel}
\author[13]{G.~Sigl}
\author[14]{A.~van Vliet} 

\affiliation[1]{Theoretical Physics IV, Faculty for Physics \& Astronomy, Ruhr University Bochum, Germany}
\affiliation[2]{Ruhr Astroparticle and Plasma Physics Center (RAPP Center), Germany}
\affiliation[3]{Institut d’Astrophysique de Paris (IAP), Sorbonne Universit\'e, Paris, France}
\affiliation[4]{Laboratoire de Physique Nucl\'eaire et de Hautes Energies, Sorbonne Université, Paris, France}
\affiliation[5]{Department of Space, Earth and Environment, Chalmers University of Technology, Gothenburg, Sweden}
\affiliation[6]{Instituto de Física Teórica UAM-CSIC, Madrid, Spain}
\affiliation[7]{Departamento de Física Teórica, Universidad Autónoma de Madrid, 
Madrid, Spain}
\affiliation[8]{Bergische Universität Wuppertal, Department of Physics, Wuppertal, Germany}
\affiliation[9]{Department of Physics, University of Oxford, Oxford, United Kingdom}
\affiliation[10]{Institute of High Technology, Immanuel Kant Baltic Federal University, Kaliningrad, Russia}
\affiliation[11]{S.~Kovalevskaya North-West Center for Math.~Research, I.~Kant Baltic Fed.~Univ., Kaliningrad, Russia}
\affiliation[12]{Faculty of Comput.~Mathematics and Cybernetics, Lomonosov Moscow State University,  Moscow, Russia}
\affiliation[13]{II.~Institute for Theoretical Physics, Universit\"at Hamburg, Hamburg, Germany}
\affiliation[14]{Department of Physics, Khalifa University of Science and Technology, Abu Dhabi, United Arab Emirates}

\abstract{We present CRPropa 3.3, the latest release of the publicly available Monte Carlo framework for simulating the propagation of high-energy particles in astrophysical environments. This version introduces significant extensions that enable multi-messenger studies across a broad energy range, from GeV to ZeV. New features include explicit time tracking, time-dependent advection fields, and support for position-dependent radiation backgrounds, for more realistic simulations of Galactic and extragalactic propagation. Nuclear cross sections have been updated and expanded up to lead (Z=82). We illustrate some of these new features, including acceleration at moving shocks and gamma-ray propagation in the interstellar radiation field. Together, these improvements establish CRPropa 3.3 as a comprehensive tool for modelling cosmic rays, gamma rays, and their secondaries in structured, time-dependent environments, setting the stage for next-generation multi-messenger astrophysics.
}

\ConferenceLogo{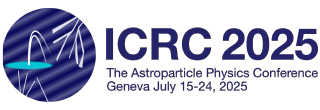}

\FullConference{%
The 39th International Cosmic Ray Conference (ICRC2025)\\
  15--24 July, 2025\\
  Geneva, Switzerland}

\begin{document}
\maketitle

\section{Introduction}

At the dawn of the multimessenger astrophysics age, there is an evident need to combine information of multiple messengers simultaneously, to interpret observations. To fully exploit the wealth of data from cosmic rays (CRs), gamma rays, and neutrinos, it is essential to model the transport and interactions of particles within their birthplaces, across cosmic distances, to Earth, all while considering their interactions and production of secondary particles over several decades in energy and time.  This requires simulation frameworks that are both physically comprehensive and computationally efficient.

CRPropa is a publicly available Monte Carlo code designed to address this challenge. It simulates high-energy particle propagation through complex magnetic fields and radiation backgrounds, accounting for a wide range of physical processes, including electromagnetic cascades and nuclear interactions. Version~3.3 features some important advancements, extending the framework's modular structure and expanding its capabilities across several cosmic domains.

In this work, we present CRPropa~3.3, an update that builds upon the foundations laid by earlier versions~\cite{CRPropa3, CRPropa32}. This release incorporates a number of technical enhancements and newly-implemented models. These include support for spatially-dependent photon fields, including the interstellar radiation field (ISRF), detailed in section~\ref{ssec:isrf}, and extended treatment of heavy and trans-iron nuclei in section~\ref{subsec:superheavy}. Technical improvements to performance and usability are summarised in section~\ref{sec:tech}. To demonstrate the utility of these features, selected simulation use cases are presented in section~\ref{sec:phys}.

\section{Technical improvements}
\label{sec:tech}

\subsection{New Galactic magnetic field models}

Two new Galactic magnetic field models have been implemented in CRPropa. The \texttt{UF23Field} class provides the eight different models by Unger \& Farrar (UF24)~\cite{UF24}; either the total coherent field can be called or --- for analysis purposes --- only individual parts, such as the spiral field. These fields are the updated versions of the well-established Jansson \& Farrar (JF12) fields~\cite{JF12}, however, a turbulent component is not yet available. So simulations should only be compared when the turbulent field is turned off in the JF12 models. Also available is now the large-scale halo field by Korochkin \textit{et al.} (KST24)~\cite{KST24}, which takes into account the Earth's location in the local bubble. Note that the Earth's position in Galactic coordinates --- usually reported in the documentation --- varies from model to model and needs to be adjusted to get accurate results.

\subsection{New EBL models}

Modeling the propagation of UHECRs from extra-galactic sources to the Milky Way requires a good knowledge of their loss processes. One of the critical targets is the extragalactic background light (EBL). Therefore, recent updated models of the EBL have been implemented in CRPropa. The models by Saldana-Lopez et al.~\cite{SL21} and Finke \textit{et al.}~\cite{F22} are available through their classes \texttt{IRB\_Saldana} and \texttt{IRB\_Finke22}, respectively. 
Note that new custom models can be implemented by the user, following the instructions in the documentation.

\subsection{Explicit time dependence}

An explicit time variable has been introduced in CRPropa, implemented as a new candidate property that is incremented at each propagation step. Currently, the limit $v = c$ still applied; however, this might be relaxed in the future, enabling the propagation of lower energy particles (below kinetic energies of $\sim\mathrm{GeV}$). Furthermore, corresponding outputs and units were added to the code.

As a first example, time-dependent advection fields have been implemented. To support this, the interface for the \texttt{Advection} base class has been modified to include  time dependence, in addition to the existing spatial dependence. In section \ref{subsec:acceleration}, acceleration at moving shocks is shown as a new physics use case.

\subsection{Other improvements}

Running CRPropa simulations on large computing facilities requires knowledge about the needed resources before the simulation. The first estimation of the needed run-time can be drawn from smaller test simulations, but for the final application one often gets to the point where the estimated time is not enough and the simulation has to be stopped. In the new CRPropa version, the state of all particles within the simulation chain can be stored and the simulation can be restarted from the point where it was stopped.

\section{New features}
\label{sec:new}

\subsection{Inclusion of nuclei up to lead}
\label{subsec:superheavy}

\begin{figure}[h]
    \centering    \includegraphics[width=0.75\linewidth]{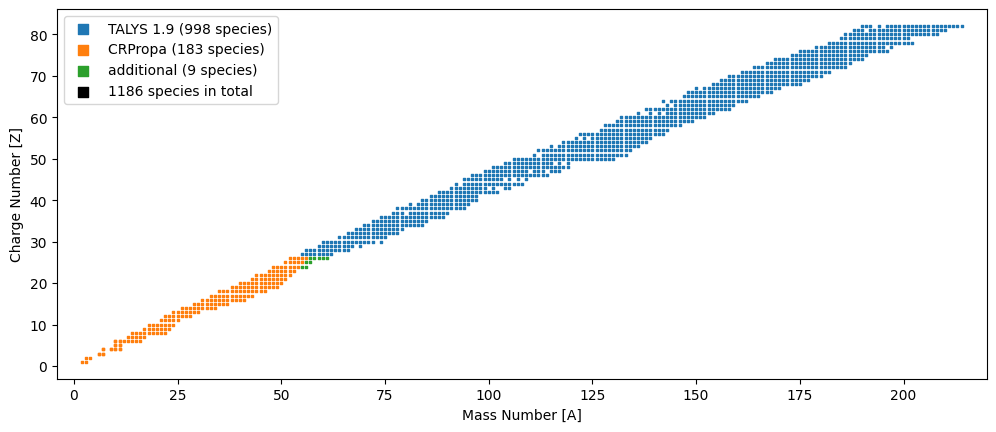}
    \caption{Nuclear species included in version 3.3, compared to the previously implemented ones.}
    \label{fig:nuclear_chart}
\end{figure}

To improve the coverage and accuracy of nuclear data for astrophysical and high-energy physics applications, additional cross sections have been included for nuclei up to lead (Z = 82). The cross sections were previously compiled~\cite{Morejon2020} and generated using TALYS~1.9 ~\cite{Koning2021Talys-1.96/2.0}. Figure~\ref{fig:nuclear_chart} illustrates the nuclear species now present in CRPropa, following this extension. The newly added nuclei (blue) increase the number of available species by more than a factor of five compared to the previous version available species (orange) which were obtained using TALYS~1.8 and other sources. In addition, some other nuclei (green) were included to fill the gap to heavier species and provide better continuity of the disintegration chains in accordance to the selection criteria.

The selection of new nuclei was guided by two main criteria: stability and completeness of decay channels.
First, only nuclei with lifetimes exceeding 100 seconds were included, a more restrictive threshold than the 1-second limit used below iron~\cite{Kampert2013}, yet sufficient to maintain coverage while significantly reducing the number of species. This corresponds to a decay length of 10~Mpc at a Lorentz factor of $10^{13}$. Second, for each nucleus, disintegration channels were retained only if they contributed at least 0.1\% to the energy-weighted average cross section, ensuring that the thinned set captured at least 90\% of the unthinned total. This reduced the number of channels from 311,403 to 16,956 (less than 6\% of the original), while preserving physical accuracy~\cite{Kampert2013}.

\begin{figure}[ht]
    \centering
    \includegraphics[width=0.52\linewidth]{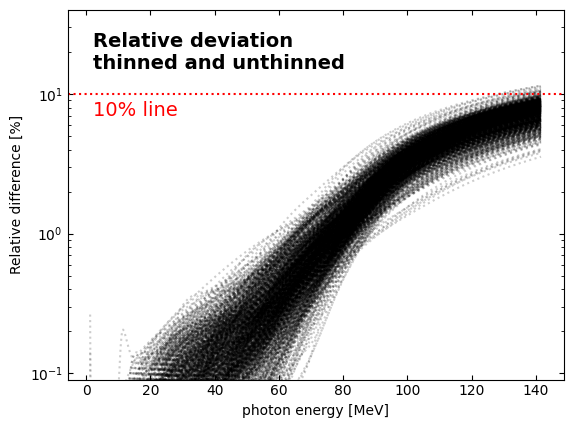}
    \caption{Relative difference between thinned and unthinned channel selection for all nuclei included. Photon energy in the rest frame of the nucleus.}
    \label{fig:thinning_error}
\end{figure}

Figure~\ref{fig:thinning_error} illustrates the relative difference between energy-weighted average cross sections calculated from the full ($\tilde \sigma_{\alpha=100\%}$) and the reduced ($\tilde \sigma_{\alpha=90\%}$) channel sets:
\begin{equation}
    \frac{\left|\tilde \sigma_{\alpha=100\%} - \tilde \sigma_{\alpha=90\%}\right|}{\tilde \sigma_{\alpha=90\%}} \,.
    \label{eq:relative_difference}
\end{equation}
Most nuclei differ by less than 10\% across the energy range, with larger deviations near the photopion threshold ($\sim 140 \; \text{MeV}$).

The energy-weighted average cross section
\begin{equation}
    \tilde \sigma(\varepsilon) = \frac{2}{\varepsilon^2} \int_0^{\varepsilon} \varepsilon' \sigma(\varepsilon') \, \text{d}\varepsilon'
    \label{eq:energy_averaged_cross_section}
\end{equation}
\noindent is independent of any specific target photon field unlike, the interaction rate $\lambda(\gamma)$
\begin{equation}
    \lambda(\gamma) = \int_0^{\infty} n(\epsilon) \tilde \sigma (2\epsilon \gamma) \, \text{d}\epsilon
    \label{eq:interaction_rates}
\end{equation}
\noindent which depends on the actual photon spectrum $n(\epsilon)$ ($\gamma$ is the Lorentz factor of the nucleus). Hence, the completeness of the decay channels holds for rates computed with any target photon spectrum.

\subsection{Position-dependent radiation field}
\label{ssec:isrf}

Various astrophysical environments are characterised by low-energy photon fields that vary spatially across different scales. One of the most important ones is the interstellar radiation field (ISRF), spanning from infrared to ultraviolet wavelengths. Several models describe its spectral intensity across the Galaxy. Such backgrounds can suppress gamma-ray and electron propagation above tens of TeV, as gamma rays may undergo pair production with ISRF photons, triggering electromagnetic cascades in our galaxy~\cite{di2025revisiting}. 

In CRPropa, the \texttt{PhotonBackground} has been implemented to host not only homogeneous radiation fields (e.g.~CMB), but also spatially dependent radiation fields, such as the ISRF. Position-dependent fields like the ISRF are specified through spatial grids of spectral intensities. From these, inverse mean free paths for the corresponding processes -- in this case, only electromagnetic interactions -- are available to the code. These tables can be computed using the standard procedure for generating \href{https://crpropa.github.io/CRPropa3/pages/example_notebooks/custom_photonfield/custom-photon-field.html}{custom photon fields} in CRPropa. The new \texttt{InteractionRates} class handles both homogeneous and spatially-varying backgrounds by locating the nearest grid node to the particle's position at each step\footnote{Nearest-node searches use the \href{https://github.com/jlblancoc/nanoflann/tree/master?tab=readme-ov-file}{nanoflann} C++ library.}. Users can also define a \texttt{Surface} to limit calculations to specific regions.

These tools enable detailed modelling of radiation environments in a range of settings, including galaxy clusters, superbubbles, etc.

\section{New physics use cases}
\label{sec:phys}

\subsection{Position-dependent radiation fields}

\begin{figure}[htb]
    \centering
    \includegraphics[width=0.6\textwidth]{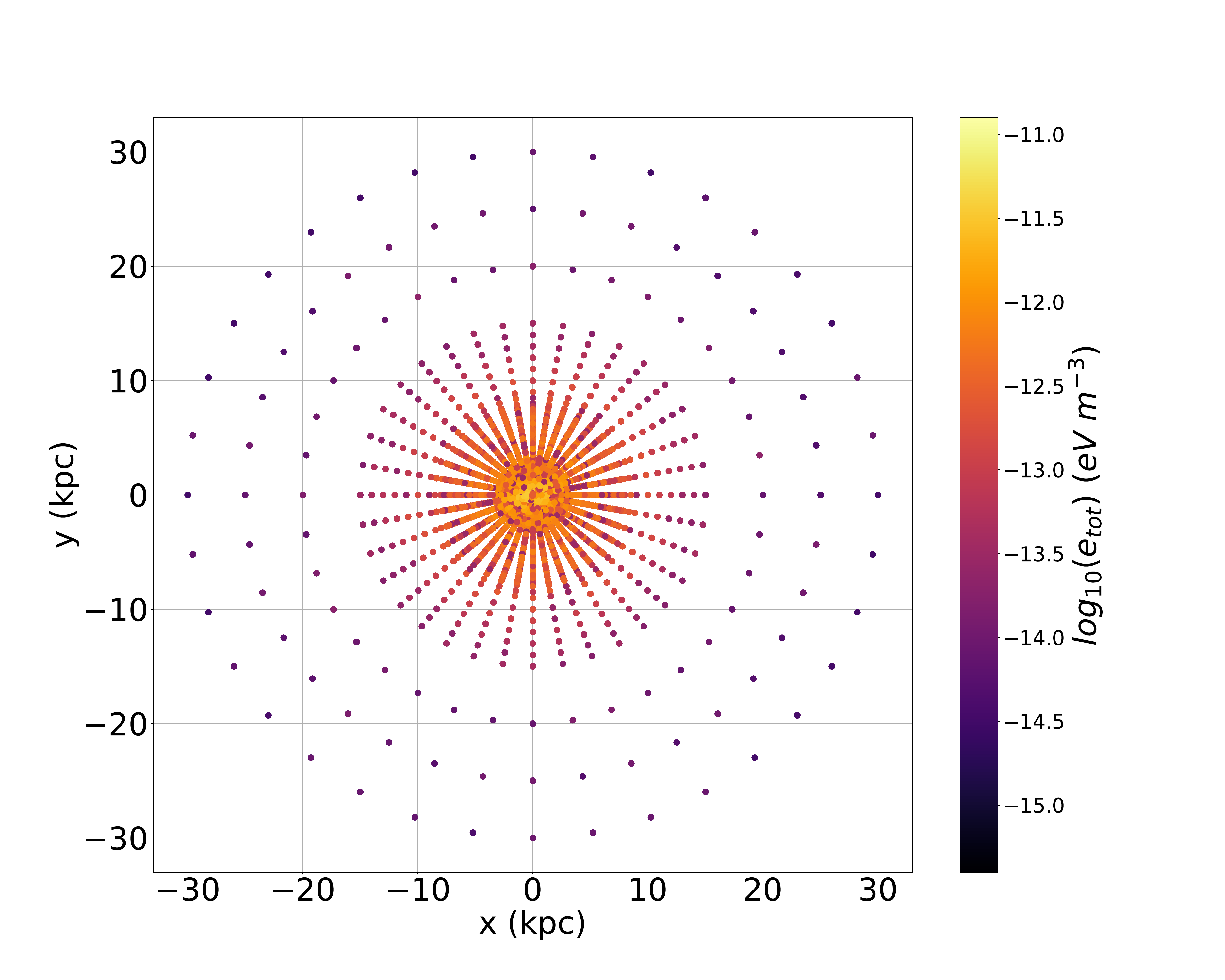}
    \caption{Nodes at the galactic plane, \textit{seen from above}, of the ISRF grid from the F98 model.  The colour bar refers to the integrated energy density of the radiation field at each node.}
    \label{fig:enDensNodeF98}
\end{figure}

\begin{figure}
    \centering
    \includegraphics[width=0.6\textwidth]{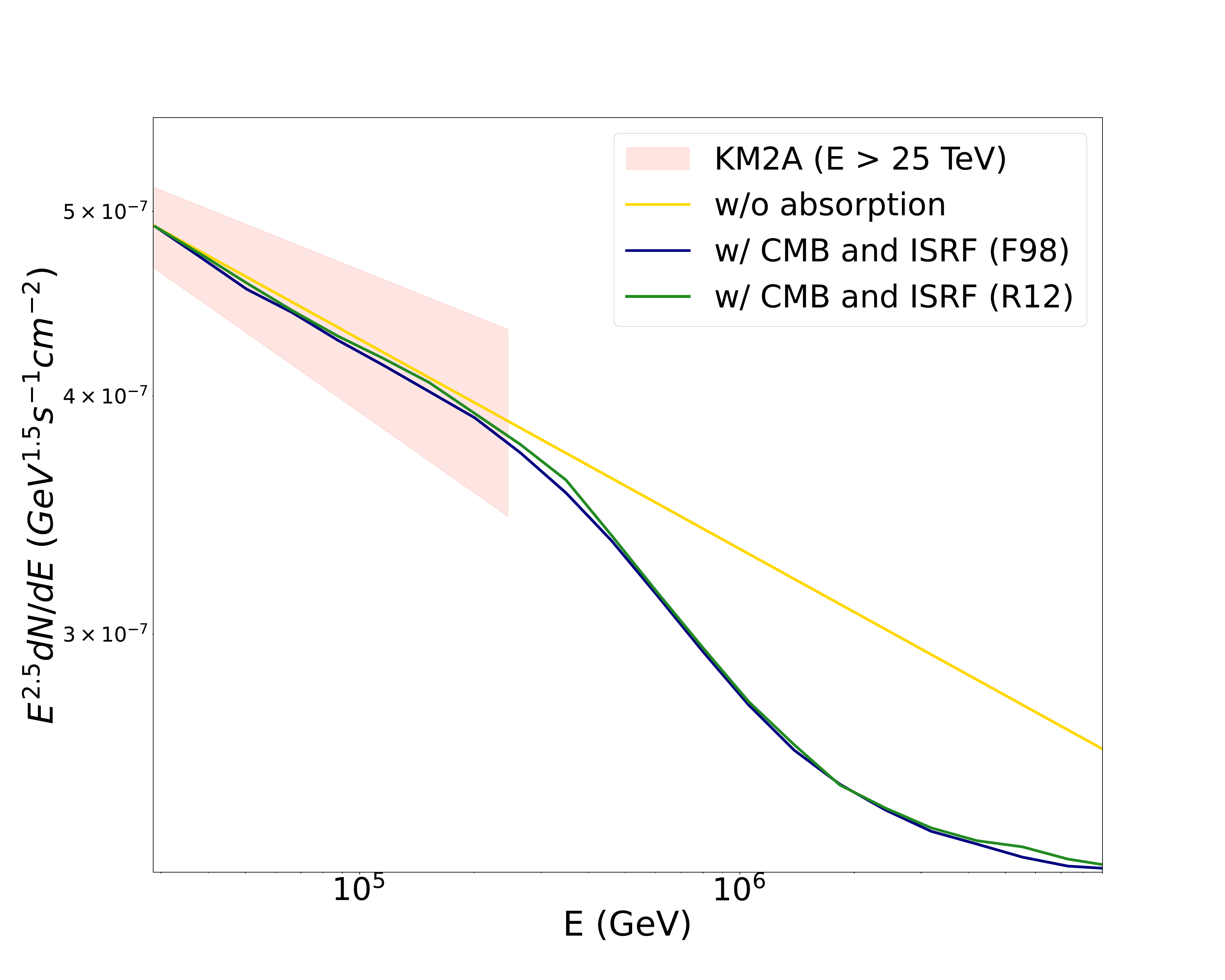}
    \caption{
    Observation of the source 1LHAASOJ1825-1256u by LHAASO-KM2A, modelled as a power law with uncertainties~\cite{cao2024first}. The simulated spectra, including Galactic photon backgrounds, are compared to a naive extrapolation without propagation effects.}
    \label{fig:LHAASOex}
\end{figure}

To interpret the mechanisms behind the recently-observed PeV emissions, the detailed treatment of gamma-ray propagation in Galactic environments is crucial. Here we present a spatially dependent implementation of the ISRF, to investigate their impact on the propagation of gamma rays in the Milky Way. The two implemented ISRF models are the ones from Refs.~\cite{freudenreich1998cobe}, hereafter F98, and~\cite{robitaille2012self}, R12, as implemented in Ref.~\cite{porter2017high}. In Fig.~\ref{fig:enDensNodeF98}, the integrated energy density grids for each node in the Galactic plane. In Fig.~\ref{fig:LHAASOex}, we report the simulated energy spectra of a Galactic source from the first LHAASO catalog~\cite{cao2024first}, 1LHAASOJ1825-1256u. 

\subsection{Acceleration at moving shocks}
\label{subsec:acceleration}

\begin{figure}[ht]
    \centering
    \includegraphics[width=0.65\linewidth]{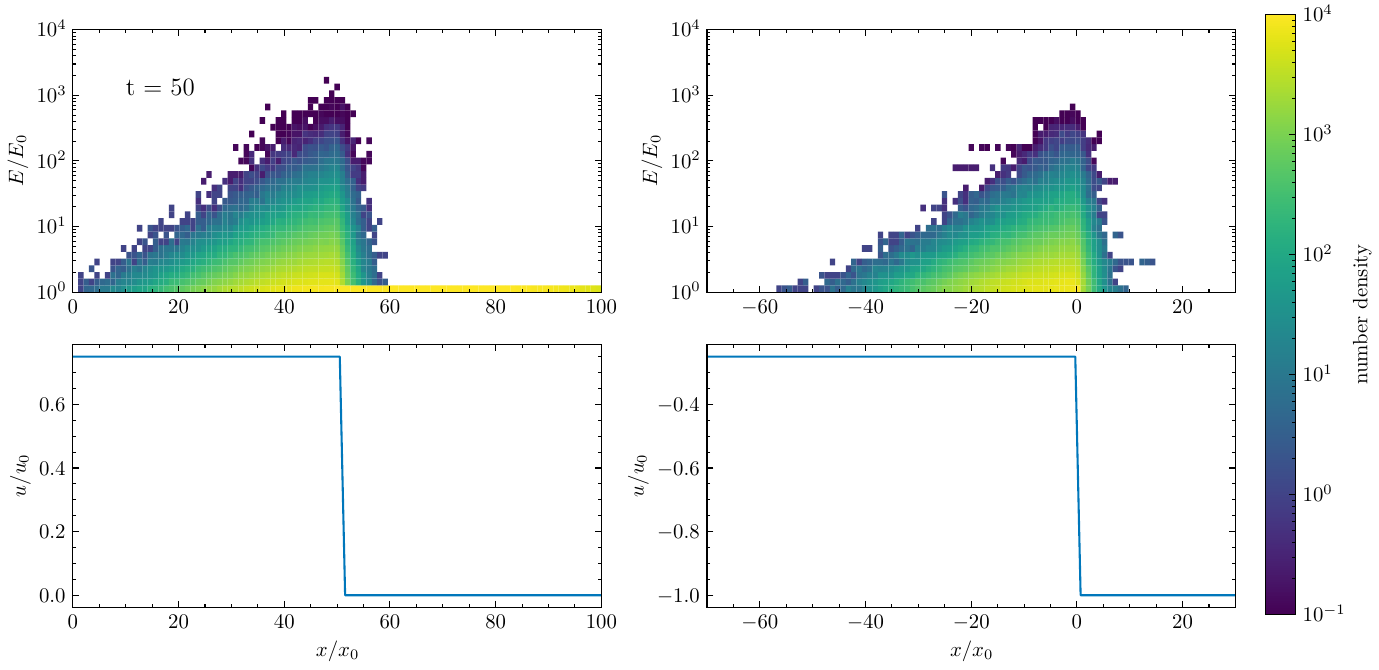}
     \includegraphics[width=0.65\linewidth]{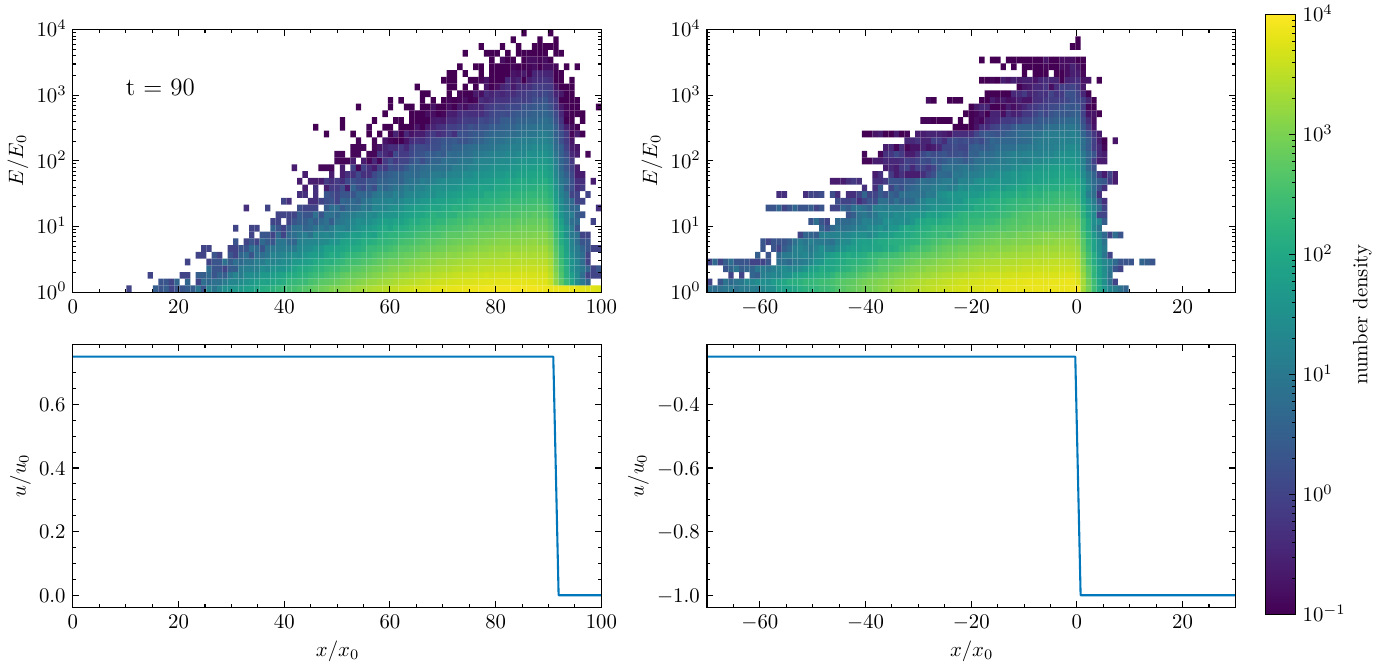}
    \caption{\textit{Left:} Shock profiles and space-energy histogram in lab frame. Pseudo-particles are injected in the undisturbed medium, the shock is moving through it. \textit{Right:} Stationary wind profile and space-energy histogram. Pseudo-particles are injected at the shock. For both columns, the top row shows spectra and wind profiles at $T=50$ and the bottom row shows them at $T=90$.}
    \label{fig:DSA-hist}
\end{figure}

We present diffusive shock acceleration (DSA) at moving shocks as the first use case for the new time-dependent advection fields. The spectra obtained at a moving shock ($v_\mathrm{sh} = 1$) are compared to those obtained in the frame in which the shock is stationary ($u_\mathrm{sh} = 0$). The advection field \texttt{OneDimensionalTimeDependentShock} is used which takes the arguments $v_\mathrm{sh}$, the velocities $v_0$ and $v_1$ as the velocities in front of the shock $v_0$ and behind the shock $v_1$. Note that the Rankine-Hugoniot conditions are only valid in the shocks frame ($u_\mathrm{sh} = 0$) and that the velocities $v_0$, $v_1$ have to be chosen such that the compression ratio $q = u_0/u_1$ matches e.g.~that of a strong shock $q = 4$, where $u_0, u_1$ denote the upstream $u_0 = v_0 - v_\mathrm{sh}$ and downstream $u_1 = v_1 - v_\mathrm{sh}$ velocities, respectively. When the velocities are not set accordingly, harder or softer spectra can emerge at the shock. Here, velocities in the lab frame are set to $v_0 = 0, v_1 = 3/4$, which corresponds to $u_0 = -v_\mathrm{sh} = -1$, $u_1 = -1/4$ in the shock frame. Lastly, \texttt{OneDimensionalTimeDependentShock} takes the shock width $l_\mathrm{sh}$; the shock profile is smoothed out with an hyperbolic tangent. For a discussion on the shock width and constraint on the diffusion coefficient/time step see, e.g.~\cite{Aerdker-etal-24}. Figure \ref{fig:DSA-hist} shows the shock profiles in lab frame $v(x)$, and in shock frame $u(x)$ and the corresponding (pseudo-)particle distributions in space and energy at two times, $t = 50, 90$. Pseudo-particles are injected with energy $E_0$ in the upstream region (lab frame), at the shock at $x = 0$ (shock frame).
In case of time-dependent advection fields, the Green's function approach to construct time-dependent solutions cannot be applied. Thus, more pseudo-particles have to be injected. While this comes at a higher computational cost, there are no uncertainties due to the time integration and "recycling" of pseudo-particles (visible as horizontal stripes in the right column of fig.~\ref{fig:DSA-hist}). A detailed discussion of uncertainties is given in \cite{MertenAerdker25}.

\subsection{External plugins}
\label{sec:plugins}

Several plugins have been developed to extend CRPropa's capabilities. While they are external to CRPropa, they can certainly be useful for specific applications. Some of them are listed below.
\begin{itemize}[noitemsep,nolistsep]
  \item \textbf{Hadronic interactions}\footnote{\url{https://gitlab.ruhr-uni-bochum.de/doernjkj/hadronic-interaction-in-crpropa}}: The plugin presented in \cite{Doerner2025} implements the interaction of CR protons with protons from the ambient gas.
  \item \textbf{Extreme electromagnetic cascades}\footnote{\href{https://github.com/GDMarco/EMMuonPairProduction}{Muon}, \href{https://github.com/GDMarco/EMElectronMuonPairProduction}{electron-muon}, \href{https://github.com/GDMarco/EMTauonPairProduction}{tauon} and \href{https://github.com/GDMarco/EMChargedPionPairProduction}{charged pion} pair productions.}: Treats the production of heavy leptons or hadrons from leptonic processes~\cite{DiMarco2025}. 
  \item \textbf{Interface with PYTHIA}\footnote{\url{https://github.com/GDMarco/CRPYTHIAxDecays}}: This additional plugin interfaces with the PYTHIA event generator~\cite{Bierlich2022} for the treatment of decays~\cite{DiMarco2025}.
  \item \textbf{Lorentz invariance violation}\footnote{\url{https://github.com/rafaelab/LIVpropa}}: Treatment of Lorentz invariance violation effects in quantum electrodynamics, following~\cite{Saveliev2024}.
\end{itemize}

\section{Summary and Outlook}
\label{sec:outlook}

In this contribution, we presented CRPropa 3.3, the latest major update to the CRPropa framework for high-energy astroparticle simulations. This release marks a significant step toward a unified, flexible, and extensible platform for multi-messenger astrophysics across a vast energy range -- from GeV to ZeV. The new features, which range from time-dependent advection and position-dependent photon fields to the inclusion of heavy nuclei up to lead, allow for more realistic simulations in various astrophysical environments. These updates not only make the simulations more realistic and more efficient, but also broaden the range of phenomena that can now be consistently modelled.

In the future, we are planning to perform further improvements, such as  the inclusion of hadronic interactions, and a more consistent and general treatment of time dependence. Moreover, continued plugin development ensures that the framework remains modular and responsive to the needs of the community. This will bring CRPropa closer to being a universal and flexible tool for Galactic and extragalactic particle propagation, while existing and future plugins guarantee an optimal level of flexibility for the user. 

\vspace{0.1cm}
\noindent\textbf{Acknowledgements}\\
{\footnotesize We thank all those who have contributed to the recent CRPropa developments, in particular: A.~Korochkin and M.~Unger for providing new Galactic magnetic field models; J.~Schmidt and I.~Vovk for adding new EBL models.}

{\footnotesize 
\noindent RAB is funded by the Agence Nationale de la Recherche (ANR), project ANR-23-CPJ1-0103-01. SA, JD, KHK, LMe, LMo, LS, and JT acknowledge financial support by the DFG (SFB\,1491). Further supported by the project \textit{Multi-messenger probe of Cosmic Ray Origins (MICRO)}, project no.\ 445990517 (LS, JT, KHK). KHK and GS acknowledge support by the BMBF Verbundforschung under grants 05A20PX1 and 05A20GU2. GS acknowledges support by the Deutsche Forschungsgemeinschaft (DFG, German Research Foundation) under Germany’s Excellence Strategy – EXC 2121 Quantum Universe – 390833306 and by the Bundesministerium für Bildung und Forschung, under grant 05A23GU3. AvV acknowledges support from Khalifa University’s internal FSU-2022-025 and RIG-2024-047 grants. GDM's work is supported by FPI Severo Ochoa PRE2022-101820 grant.}

\setlength{\bibsep}{0pt plus 0.3ex}
\setstretch{0.9}

\end{document}